# Defect complex formation in TiO$_2$ nanoparticles by sub-band excitation

Neha Luhakhra and Sanjiv Kumar Tiwari[*]

*Department of Physics and Materials Science, Jaypee University of Information Technology Waknaghat Solan, H.P 173234, India*

Anatase TiO$_2$ is an indirect band-gap material with large degree of sub-stoichiometry, selective excitation of its native defect reveals polychromatic emission in blue, green and red spectral regions due to self trapped excitons and singlet-singlet transition of defect states. Electron spin resonance measurement in dark shows presence of isolated Ti$^{+3}$, deep lying, and surface lying V$_O$ at 1.986, 2.04 and 2.023 g values respectively, whereas, on exposure with visible light in situ ESR measurement reveals the formation of defect complex of Ti$^{+3}$-V$_O$ and Ti$^{+4}$-V$^+_O$.

**Key words**: Defects, luminescence, Optical materials and properties, nanocrystalline materials

[*] Corresponding author.

Email address: sanjivkumar.tiwari@juit.ac.in

1. **Introduction**

Titanium dioxide $TiO_2$ crystallizes in three different structures commonly named as rutile ( $D^{14}_{4h}$-$P4_2/mnm$), anatae ( $D^{19}_{4h}$-$I4_1/amd$) and brookite ( $D^{15}_{2h}$-$Pbca$). Due to large degree of sub-stoichiometry, intrinsic defects of $TiO_2$ ( Ti interstitial $Ti_i$, oxygen vacancy $V_O$ and Ti vacancies $V_{Ti}$) plays crucial role in photo-physical process and energy conservation scheme . The key factor affecting the photo-physical process is either the addition of excess electron in the conduction band or the trapping of electrons at defect sites. It has been reported that in rutile the excess electron is trapped in $Ti^{+3}$ site, and $Ti^{+3}$ relax out word by 2%-4% of equilibrium bond length, whereas, in anatase the crystal remains unperturbed and excess electron is homogeneously distributed in the crystal [1-3]. However, interaction between surface defects and bulk defects or defect complex formation has not been studied yet. In this letter we report, optical emission due to defects on excitation with a photon of energy less than band gap i.e 3.1 eV and 2.75 eV and the defect complex formation on exposure with visible light by in situ ESR measurement.

2. **Experimental details**

$TiO_2$ nano-particles of size 40 nm, was purchased from SRL and used without any further purification. The UV – Visible (UV - Vis) spectra were recorded using a spectrophotometer (Perkin Elmer λ-750) and defect-related optical emission was recorded using a fluorescence spectrometer (Perkin-Elmer LS-55). Intrinsic charged state of defects and defect complexes were studied using x-band ESR measurement (BRUKER BIOSPIN, Germany, model No: EMXmicro A200-9.5/12/S/W 9.4 GHz and 2 mW power) at room temperature.

3. **Results and discussions**

Figure 1(a) and (b) shows the PL spectra of $TiO_2$ nano-particles on excitation with photon of energy 3.1 eV and 2.7 eV respectively, our electronic state calculation reveals indirect band nature of $TiO_2$ nano-particles , figure S1(b) of supplementary information, inset of figure 1(a) shows Tauc plot , which gives optical band gap of 3.2 eV. Our sample does not show any absorption peak in the visible region; however, a broad tail due to presence $V_O$ and adsorbed molecular oxygen ($O_2^-$) is evident. Sekiya et.al, had reported polarization-dependent absorption band due to $V_O$ at 435 nm and 413 nm in parallel and perpendicular configuration respectively [3]. PL spectra shows five different emission peaks at 2.72 eV, 2.54 eV, 2.34 eV, 2.20 eV and 2.09 eV respectively, corresponding peaks are shown with a dotted line and resultant of de-convolution is shown with solid lines. PL peaks at 2.72 eV and 2.54 eV were assigned due to localized and delocalized self trapped excitons (STE) respectively [4]. This assignment is in accordance with our density of state calculation also, which reveals that d orbital's of Ti atoms are localized as compared to oxygen P orbital of valance band ( figure S1(c) of supplementary information). PL peaks at 2.34 eV and 2.20 eV were attributed due to deep-lying $V_O$ and its complexes with Ti ions, whereas, PL peak at 2.09 eV were attributed due to $V_{OS}$. To distinguish the emission from STE, same sample was excited with much lower energy photon (2.7 eV) and the resulting spectra are shown in figure 1(b). ESR measurement reveals that Ti and $V_O$ both are in an ionized state; these defects are dynamic when exposed to light and form defect complexes with Ti ions. Electrons trapped at $V_O$ are mainly localized on two neighbouring Ti ions ,and association of $V_O$ with $Ti^{+3}$ gives origin of defect complex, which can be expressed as $V_O+2e^- =F_O$, $F_O+Ti^{+4}=F^++Ti^{+3}$, $V_O+e^- = F^+$ and $Ti^{+4}+e^-=Ti^{+3}$. Further, when these ionized defects selectively excited by a photon of sub-band gap energy then inter atomic excitation ( singlet to singlet) and the internal crossing( singlet to triplet) takes place, hence, transition from singlet to singlet and excited triplet state to ground state give emission in the green region, schematic of these transitions are shown in the inset of figure 1(b) which indicates singlet to singlet transition of $V_O$ (2.33 eV) and $Ti^{+3}$-$V_O$ and $Ti^{+4}$-$V^+_O$ (2.04 and 1.91 eV) . Figure 2a and 2b shows ESR spectra of $TiO_2$ in dark , and when exposed with visible light respectively. In dark, we observed ESR signal corresponding to isolated $Ti^{+3}$ at g= 1.986, deep-lying $V_O$ and surface $V_{OS}$ at g= 2.04 and 2.023, adsorbed $O_2^-$ at g= 2.15 respectively, which is in accordance with Mishra et.al [5]. With visible light exposure, signal corresponding to $Ti^{+3}$ and $O_2^-$ drastically shifts from g= 1.986 to 1.925, and g= 2.15 to 2.20 respectively. This indicates the change in the crystal environment around $Ti^{+3}$, and $Ti^{+3}$-$V_O$ defect complex formation. This observation is analogous to Yang et.al that the ESR signals at g=1.93 is due to the complex of $Ti^{+4}$, because the trapped photo-excited electron at $V_O$ ( $e^-+V_O=V^+_O$)  is shared with neighbouring Ti ions, which causes tetragonal elongation of the unit cell by columbic interaction, and degenerate $d_{xy}$, $d_{yd}$ orbital's respectively this leads to the  red shift in g value of isolated $Ti^{+3}$ and formation of $Ti^{+3}$-$V_O$ or $Ti^{+4}$-$V_O^+$ complex , whereas,  the hole is captured by adsorbed oxygen ($O_2^-+h^+=O^-$) having an unpaired electron hosted in P orbital's which induces the transition of $O^-$ ions from axial symmetry to rhombic symmetry, as a result, g value get shifted from g= 2.150 to 2.200 . The same type of observation has been reported by Chiesa et.al on the surface of polycrystalline MgO [6,7]. Origin of the signal at g=2.830 is not known, although we strongly believe that it could be due to complexes of Ti vacancies.

4. **Conclusions**

Sub-band photo excitation of nano-crystalline $TiO_2$ does not produce electrons or holes in conduction and valance band rather atomic excitation of oxygen vacancy takes place. On absorption of photons of sufficient energy $V_O$ form complexes with Ti ions and get excited to their respective excited singlet state. Radiative recombination from excited singlet state to ground singlet state gives luminescence in the visible region. These defect complexes may have wide application in photodynamic therapy and photo-catalytic activity of $TiO_2$.

**Acknowledgement**

The author would like to thanks the institute instrumentation centre at Indian Institute of Technology, Roorkee India, for ESR measurement.

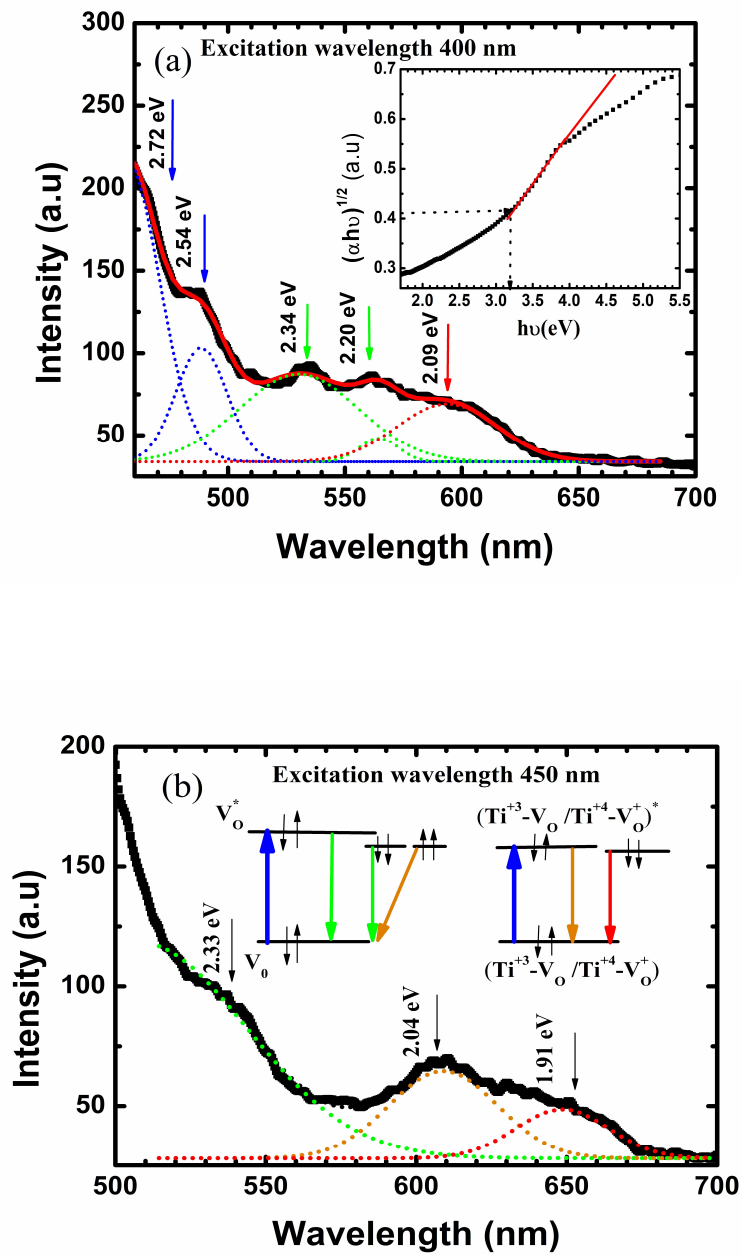

Figure1. Photoluminescece spectra of $TiO_2$ by sub band excitation by 400 nm, inset represent Tauc plot for indirect band gap material, dotted squared region in inset shows broad tail in visible region of electromagnetic spectrum, (b) Photoluminscence spectra by sub band excitation with 450 nm, PL positions are marked with arrow, PL profile was Gaussian fitted with correctness factor of $r^2=0.999$, inset represent electronic and ionic state of defect and therir complexes

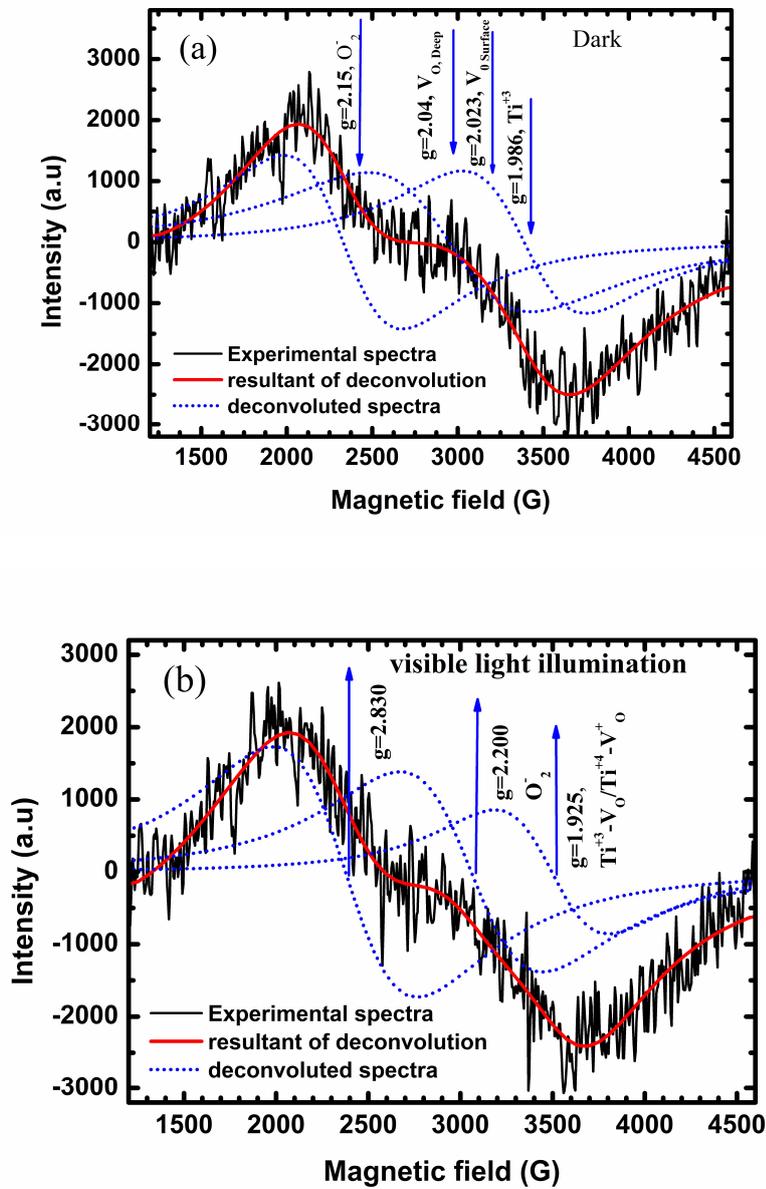

Figure 2. X-band (9.45 GHz) EPR spectra of $TiO_2$ nanoparticles, each spectra is deconvoluted with Lorentzian line profile, and resultant of de-convolution is shown with red line (a) EPR spectra in dark (b) EPR spectra with vigible light exposure, corresponding g values and their respective origins are marked with arrows.